\documentclass[prl,twocolumn,showpacs,preprintnumbers,amsmath,amssymb]{revtex4}

\usepackage[dvips]{graphicx}

\begin{document}

\title{Recombination of three ultracold fermionic atoms}

\author{H. Suno$^1$, B.D. Esry$^1$, and Chris H. Greene$^2$}
\affiliation{$^1$Department of Physics, Kansas State University, 
         Manhattan, Kansas 66506}
\affiliation{$^2$Department of Physics and JILA, University of Colorado, 
         Boulder, Colorado 80309}
\date{\today}

\begin{abstract}
Three-body recombination of identical, spin-polarized fermionic atoms 
in the ultracold limit is investigated. The mechanisms for recombination 
are described in terms of the ``scattering volume'' $V_p$ in the framework 
of the adiabatic hyperspherical representation. We have 
calculated numerically the recombination rate $K_3$ as a function of 
$V_p$ and have found that $K_3$ scales as $|V_p|^{8/3}$ for small $|V_p|$. 
A comparison with experimental data is also presented.
\end{abstract}


\maketitle

Recently, the quantum degenerate regime was attained in ultracold gases 
of fermionic atoms such as $\rm{^{40}K}$~\cite{DeMarco} and 
$\rm{^6Li}$~\cite{OHara,Truscott}. Part of the motivation for these 
experiments is to observe a pairing of fermions, leading 
to a superfluid state. One important factor limiting the achievable density 
in these degenerate Fermi gases (DFG's) of trapped atoms is 
the loss of atoms through three-body recombination.  Such losses occur 
when three atoms scatter to form a molecular bound state and a third atom ---
$\rm K+K+K\rightarrow K_2+K$, for instance. 
The kinetic energy of the final state particles causes them to escape 
from the trapping potential.

While ultracold three-body recombination of identical, spin-polarized bosons has been
theoretically studied because of its importance for Bose-Einstein 
condensates, recombination of identical fermions has not yet been considered. 
For bosons, Fedichev {\it et al.}~\cite{Fedichev} predicted that 
the recombination rate $K_3$ grows with the two-body $s$-wave scattering 
length $a_s$, namely $K_3\propto a_s^4$, for $a_s>0$. This scaling was later confirmed 
by Nielsen and Macek~\cite{Nielsen} who also pointed out that it should hold for
negative $a_s$. The $a_s^4$ scaling law for 
{\em both} signs of $a_s$ was indeed obtained by Esry {\it et al.}~\cite{Esry}, 
Bedaque {\it et al.}~\cite{Bedaque}, and Braaten and Hammer~\cite{Braaten}. 

In the case of fermions, however, the Pauli exclusion principle prohibits 
$s$-wave scattering of atoms in identical spin states, thus leaving 
only $p$-wave collisions.  The relevant low-energy scattering 
parameter in this case is the two-body $p$-wave 
``scattering volume'' defined as
\begin{equation}
V_p = - \lim_{k\rightarrow 0} \frac{\tan\delta_p(k)}{k^3}, 
\label{ScatteringVolume}
\end{equation}
where $\delta_p(k)$ is the $p$-wave scattering phase shift and $k$ is 
the wave number. The scattering volume $V_p$ is related to the 
$p$-wave scattering length $a_p$ (see, for instance, Ref.~\cite{Bohn}) by $V_p=a_p^3$. 
We choose $V_p$, rather than $a_p$, as the parameter to characterize the 
three-body recombination of fermions since an artificial 
nonanalyticity is introduced into $a_p$ when taking the cube root of the 
quantity in the right-hand side of Eq.~(\ref{ScatteringVolume}). 

Even though recombination of identical fermions is suppressed at ultracold
temperatures by the Pauli principle, it does not vanish.  In fact, it has been shown
that the rate is proportional to $E^2$ at low collision energies~\cite{ThresholdPaper}.
While this rate remains negligible under typical experimental conditions,
it can become substantial near a Feshbach resonance.   The $E^2$ threshold law
no longer applies, and the recombination rate tends to the limit imposed by unitarity ---
often comparable to or larger than the rates for boson systems.  Feshbach resonances
are, of course, extremely useful tools for the experimentalist, so understanding
the behavior near such a resonance is crucial.  So far, such resonances 
have been observed, for example, in systems of $\rm{^{40}K}$~\cite{Loftus}
and $\rm ^6Li$~\cite{OHara}. This 
resonant tuning was also observed  for other alkali species 
in BEC experiments~\cite{Inouye}. 

This Letter treats the three-body recombination of 
identical, spin-polarized fermions in the ultracold limit. We examine 
the recombination rate $K_3$ as a function of the 
scattering volume $V_p$ by numerically solving the three-body scattering
problem. First of all, since the atoms are spin polarized and thus in a completely
symmetric spin state, the spatial wave function must be completely antisymmetric
in order to satisfy the Pauli principle.  This fact, combined with
the generalization of Wigner's threshold law to $K_3$, shows that
the $J^\Pi=1^+$ symmetry dominates at threshold, where $J$ is the total orbital
angular momentum and $\Pi$ is the parity of the system.  It follows that the
recombination rate depends on the collision energy $E$ as
$E^2$ near threshold~\cite{ThresholdPaper}.  
(The same
analysis applied to the boson case yields the familiar result that the 
recombination rate is constant at threshold due to the $0^+$ symmetry.)
Therefore, we consider only the $1^+$ case.  Dimensional analysis, together 
with this $E^2$ law, suggests a $|V_p|^{8/3}$ scaling of the recombination rate $K_3$ 
as opposed to the $a_s^4$ scaling for bosons. 

The interaction potential used is a sum of triplet two-body potentials, 
i.e. $V=v(r_{12})+v(r_{23})+v(r_{31})$. This choice is appropriate 
for fully spin-polarized atoms that collide in a quartet electronic 
state. For simplicity, we model the two-body potential as either
\begin{equation}
v(r_{ij}) = D{\rm sech}^2\left(\frac{r_{ij}}{r_0}\right)
\label{sech2}
\end{equation}
or 
\begin{equation}
v(r_{ij}) = \frac{D}{1+\left(\frac{r_{ij}}{r_0}\right)^6}. 
\label{C6}
\end{equation}
The former potential has proven convenient in recombination 
calculations while the latter has a more physical van der Waal's 
tail with $C_6=r_0^6 D$.  Ideally, the results will not depend
on the particular potential used in the ultracold limit.
The parameter $r_0$ controls the
range of the potential.  The coefficient $D$, 
representing the potential depth, is treated
as an adjustable parameter that permits us to control the scattering 
volume $V_p$, thus mimicking the tuning ability of an external magnetic 
field. This coefficient also economizes the calculations substantially
since it allows us to reduce the number of two-body bound states without
sacrificing the ultracold physics.

The two-body $p$-wave scattering volume $V_p$ behaves much like the $s$-wave scattering length
as a function of the parameter $D$, displaying a tangent-like structure.
As $D$ becomes more negative and the potential becomes more attractive, 
the scattering volume passes through a pole and changes sign each time the 
potential becomes deep enough to support one additional $p$-wave bound state. 
For simplicity, we consider the parameter range for which there exists 
only one two-body bound state.

The details of the theoretical methods we employ are more completely discussed in Ref.~\cite{Suno},
so only a brief outline is given here.
We use the adiabatic hyperspherical representation with modified Smith-Whitten hyperspherical 
coordinates.  Simply put, hyperspherical coordinates 
transform the six relative Cartesian coordinates
in the center of mass frame to a set with a single length coordinate, the hyperradius $R$,
and five hyperangles.  The hyperradius can thus be thought of as 
characterizing the overall size of the three-body system.
These coordinates also allow us to easily impose the correct permutation symmetry
on the wave functions.  Solution of the adiabatic equation yields
adiabatic hyperspherical potential curves and channel functions. The 
coupled hyperradial equations are then
solved using an $R$-matrix propagation method.

\begin{figure}
\centerline{\includegraphics[scale=0.40,clip=true]{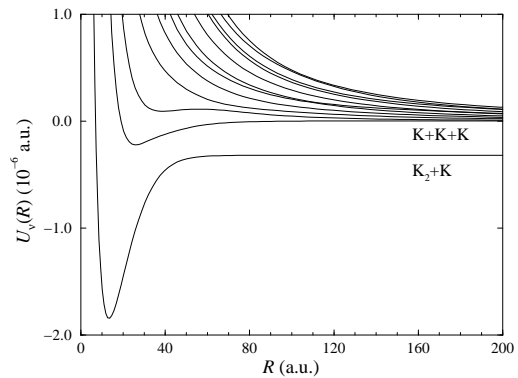}}
\caption{The lowest 12 adiabatic hypersperical potential curves for three spin-polarized
fermions with $V_p=-1.25\times 10^5$~a.u.$^3$ using the sech$^2$ potential with $r_0$=15~a.u.
}
\label{Figure2}
\end{figure}

The adiabatic hyperspherical representation reduces the collision of the three 
atoms to dynamics on a set of coupled hyperradial adiabatic
potentials.  These potentials bear a strong resemblance to standard molecular
Born-Oppenheimer potentials (see Fig.~\ref{Figure2}) and can be interpreted in
much the same way.  The lowest potential curve in the figure, for instance, correlates
to a bound molecule and a free atom far away.  Because the atoms are 
identical fermions, the molecule has unit angular momentum and the potential
has a centrifugal barrier.  Further, because
we consider only the $1^+$ symmetry due to its dominance at threshold, the
free atom must also be in a $p$-wave relative to the center of mass of the
molecule.  
All of the other curves in Fig.~\ref{Figure2} correlate to three free atoms.
In fact, there are an infinite number of potential 
curves associated with three-body continuum channels and approach
the three-body breakup threshold $U=0$ asymptotically.  

Three-body recombination can be seen as a transition from one three-body continuum 
channel to the recombination channel, driven by nonadiabatic coupling.  Qualitatively, the
adiabatic potentials shown in Fig.~\ref{Figure2} display analogous behavior, as functions of $V_p$, 
as the adiabatic potentials for bosons show as functions of $a_s$~\cite{EsryJPB,Esry}
For both systems, the
entrance channel goes from being strongly repulsive for positive $V_p$ and $a_s$
to having an attractive well behind a potential barrier for negative values.
The entrance channel for the fermions is generally more repulsive, however, than the
addition of a simple $J$=1 centrifugal potential term to the boson curves would give
due to permutation symmetry considerations~\cite{ThresholdPaper}.  The recombination
channels are also very similar, although it should be noted that there is no Efimov
effect~\cite{efim70} for fermions in the limit $|V_p|\rightarrow\infty$.  Efimov physics 
plays a key role in the interpretation of the ultracold recombination of bosons~\cite{Esry}.

The primary difference between fermion and boson systems lies in the nonadiabatic
coupling.  While it is still similar for negative $V_p$ and $a_s$, for positive
values the similarities end.  Where the coupling strength for bosons shows a definite peak
whose position increases linearly in $a_s$, the coupling strength for fermions remains
peaked at small $R$ with a slowly decaying shoulder whose extent grows in proportion
to $V_p^{1/3}$.  For negative values of $V_p$, we thus expect that the fermion recombination
rate can show resonant enhancement due to three-body shape resonances just as for bosons,
but we do not expect an infinite series of such resonances since this was a consequence
of the Efimov effect.  We have not, however, seen evidence of such resonances 
as $V_p\rightarrow -\infty$ in our
calculations.  

To calculate the three-body recombination rate,  we solve the hyperradial equations~\cite{Suno}
for potentials like those shown in Fig.~\ref{Figure2}.  The numerical hyperradial
wave functions are matched to the appropriate Bessel functions based upon
the known form of the long-range potentials~\cite{Suno}.  The $S$-matrix is 
obtained and the generalized cross section for three-body recombination is calculated:
\begin{equation}
\sigma = \frac{192(2J+1)\pi^2}{k^5}\sum_{f,\lambda}|S_{f,\lambda}|^2=\frac{576\pi^2}{k^5}\sum_{f,\lambda}|S_{f,\lambda}|^2.
\label{CrossSection}
\end{equation}
Here, $k=\sqrt{2\mu E/\hbar^2}$ is the hyperspherical wave number 
in the incident three-body continuum channel, and the indices
$\lambda$ and $f$ label initial three-body 
continuum and final recombination channels, respectively. 
As it turns out, the numerical prefactor in Eq.~(\ref{CrossSection}) 
--- which is determined by permutation symmetry ---
is the same as for three identical bosons~\cite{Esry} (recall that we
are considering the $J^\pi=1^+$ symmetry).  The {\it event}
rate constant per atomic triad is then defined simply as $K_3 =\frac{\hbar k}{\mu}\sigma$.
This quantity is related to the {\it atom-loss} rate constant $L_3$ by 
$L_3=3K_3/6$~\cite{Esry}.

Because the rate depends strongly on the collision energy in the ultracold
regime, it must be thermally averaged in order to compare with experimental data.
Following Ref.~\cite{BurkeThesis}, we have derived the thermally averaged 
recombination rate constant to be
\begin{eqnarray}
\langle K_3\rangle(T) &=&\frac{\int K_3(E) E^2e^{-E/k_BT}dE}
{\int E^2 e^{-E/k_BT}dE} \nonumber \\
&=&\frac{2}{(k_B T)^3} \int K_3(E) E^2e^{-E/k_BT}dE.
\label{ThermAvg}
\end{eqnarray}

\begin{figure}
\centerline{\includegraphics[scale=0.40]{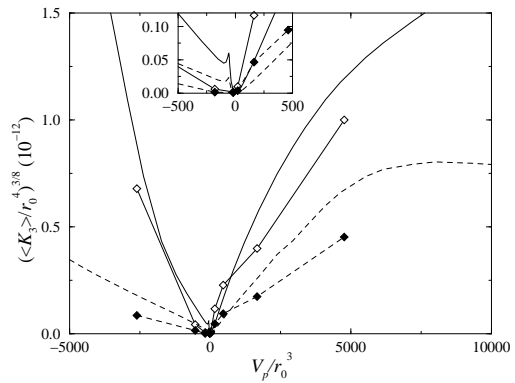}}
\caption{The recombination rate $K_3$ as a function of $V_p$ for potentials from
Eq.~(\ref{sech2}) (no symbols) and Eq.~(\ref{C6}) (diamonds).  
The solid lines corresponds to the thermally averaged rate at $T$=2~$\mu$K;
and the dashed lines, to the energy-dependent rate at $E$=2~$\mu$K.  All curves were
calculated with $r_0=15$~a.u.  The inset 
focuses on the region with small $|V_p|$ showing a St\"uckelberg oscillation.}
\label{Figure3}
\end{figure}

We show in Fig.~\ref{Figure3} $\langle K_3 \rangle$ as a function of $V_p$
for a temperature of 2~$\mu$K.  Specifically, Fig.~\ref{Figure3}
shows $(K_3/r_0^4)^{3/8}$ versus $V_p/r_0^3$.  
We plot $K_3^{3/8}$ to best reveal whether the predicted
$|V_p|^{8/3}$ scaling actually holds for the calculated rate; 
the factors of $r_0$ were included based on
general arguments of length scale invariance of the 
Schr\"odinger equation.  Because the two-body potentials in
Eqs.~(\ref{sech2})--(\ref{C6}) depend only on the combination $r_{ij}/r_0$,
$r_0$ can be defined to be the new length scale for both the two-body and
three-body systems.  This new length scale 
implies multiplying energies and temperatures by $r_0^2$, lengths by $r_0^{-1}$, and the
rate by $r_0^{-4}$.  The energy parameter $D$ in our model --- which controls $V_p$ ---
is independent of the length scale, and so its scaled counterpart, $r_0^2 D$,
can be used to control $V_p/r_0^3$.  Having calculated the rate as
a function of $V_p$ and energy for a fixed $r_0$, we can thus obtain
the rate for any other value of $r_0$ by means of the above scaling.
The key is that at a fixed laboratory collision energy (or temperature), changing
$r_0$ effectively changes the range of the two-body potential, and the
new rate corresponds to the rate at the scaled energy $r_0^2 E$.  One
has to be careful, of course, to track the simultaneous change in the 
scaled $V_p$.

The solid lines
in the figure show the thermally averaged rates $\langle K_3 \rangle (T)$, while
the dashed lines show the energy-dependent rates $K_3(E)$.  
The symbols for the sech$^2$ potential, Eq.~(\ref{sech2}), are omitted
for clarity since the rate was obtained on a relatively dense set of $V_p$ values.
The diamonds denote the rates resulting from the second potential, Eq.~(\ref{C6}).
The rates from the two potentials --- which are qualitatively very different ---
show reasonable quantitative agreement, especially for the more experimentally
relevant thermally averaged rate.  The poorest agreement occurs at small $|V_p|$
where the rates are also small.  If, however, the rate curve for either potential
with a different value of $r_0$ and the same lab energy $E$ were plotted here,
it would look systematically different, although qualitatively similar.  
This apparent dependence on the two-body potential would seem to argue against
a ``universal'' curve as was found for bosons~\cite{Esry,Nielsen,Bedaque,Braaten}.
Based on the above length scaling arguments, though, it is clear that the scaled
rate curves --- at the same scaled energy --- are what should be compared between
different potentials and would thus come closest to a universal curve.

Figure~\ref{Figure3} shows that the $|V_p|^{8/3}$ scaling does indeed hold quite well for
$|V_p|/r_0^3$ less than roughly 2500 for $K_3(E)$, but for $\langle K_3\rangle(T)$ it
only holds over the region $|V_p|/r_0^3$ below approximately 800.  Since this scaling
is based on the threshold behavior of the rate, it is not surprising that it breaks
down sooner for the thermally averaged rate given that the averaging procedure 
includes rates from higher energies.  The fact that the scaling breaks down for
$K_3(E)$ is an indication that the fixed collision energy is no longer in the
threshold regime.

For large positive scattering volumes $V_p$, the collision energy eventually
becomes large compared to the dimer binding energy, and
$K_3$ increases more slowly than the $|V_p|^{8/3}$ scaling law.  Recalling that
for negative $V_p$ there is a barrier in the initial adiabatic hyperspherical potential,
the breakdown in the scaling law comes when the barrier sinks below the collision energy 
at large negative $V_p$.  The fact that
the recombination rate increases more rapidly than the scaling law 
with increasing $|V_p|$ simply reflects the more ready tunneling through the
barrier.  The rate continues to increase until it reaches the limit imposed
by unitarity, namely (in SI units)
\begin{equation}
K_3^{\rm max}=\frac{\hbar}{\mu}\frac{576\pi^2}{k^4}
= \frac{\hbar^5}{m^3}\frac{144\sqrt{3}\pi^2}{E^2}.
\end{equation}
This value is obtained from Eq.~(\ref{CrossSection}) by assuming that
the recombination predominantly comes from the lowest continuum channel
with unit probability.  We expect that this limit will be reached 
when $|V_p|$ becomes large enough that the barrier in the 
three-body entrance channel is lower than the collision energy.
For the conditions of Fig.~\ref{Figure3}, this value of $V_p$ is well
off the scale of the plot.

Interestingly, the inset of Fig.~\ref{Figure3} shows a 
small peak in the rate located at about $V_p/r_0^3=-60$, i.e. at B=200.8 G.
This peak is a constructive interference between two indistinguishable pathways ---
a St\"uckelberg oscillation.  This peak disappears and reappears as a function
of collision energy since the phase difference between the two pathways depends
on the energy.

\begin{figure}
\centerline{\includegraphics[scale=0.40]{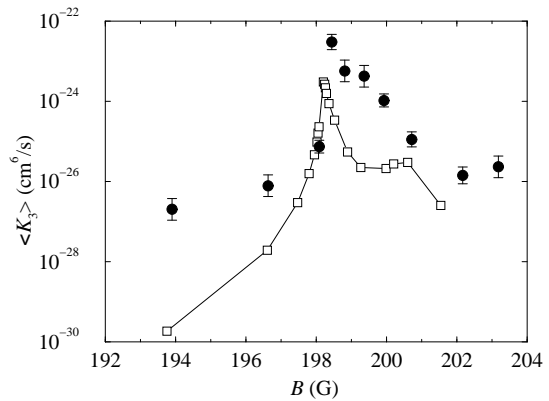}}
\caption{
The theoretical rate (solid line) and the rate measured by C. Regal {\it et al.} 
(filled circles) are plotted 
as functions of the magnetic field strength $B$.}
\label{Figure4}
\end{figure}

In Fig.~\ref{Figure4}, we compare our calculations with experimental measurements of the 
recombination rate of three ${\rm ^{40}K}$ atoms all in the $|9/2,-7/2\rangle$ 
spin state near a $p$-wave two-body Feshbach resonance.  The theoretical rates 
are calculated at $T=2.5~\mu$K, while the experimental temperature 
varies between 2 and 3~$\mu$K. The figure shows the rates
as functions of the magnetic field strength $B$.
The Feshbach resonance lies at $B_0=198.5$~G; $V_p$ is positive for $B$ below the 
resonance and negative for $B$ above.  Treating $r_0$ as a fitting
parameter, we obtain the solid line shown in the figure.  The agreement
on the rise of the resonance ($V_p>$0) is better than on the fall ($V_p<$0).
The disagreement away from the resonance is likely due to the breakdown of the
experimental assumption of purely three-body loss in determining the rate coefficient.
Near the resonance, however, our theoretical rates between one and two orders
of magnitude lower.  In fact, for all 
values of $r_0$, the theoretical values lie below the experimental ones.
This discrepancy is not surprising given that we included only a single 
partial wave ($1^+$) in our calculation.  While this symmetry controls the limiting 
behavior at threshold, it is not necessarily dominant near the resonant 
peak. In fact, limited explorations have demonstrated that other 
$J^\pi$ can be comparable to $1^+$ at the resonance peak.  The contributions
from these higher partial waves accumulate quickly since thermal
averaging, from Eq.~(\ref{ThermAvg}), gives an extra factor of roughly $(J+2)!$.

In summary, we have investigated ultracold three-body recombination of 
identical spin-polarized fermions and have numerically calculated the 
recombination rate as a function of the scattering volume.  This is the
first such study that we are aware of, and much remains to be understood
about these fermion systems.  Moreover, with some slight
modifications to our numerical method, it is possible to treat 
systems with only two identical particles.
Three-body recombination of such systems plays an important 
role in experiments on mixed-spin state Fermi gases and mixed 
Bose-Fermi gases.

This work was supported in part by the National Science Foundation and
the Research Corporation. We thank C. Ticknor and J. Bohn for sharing their 
two-body theoretical scattering data.
We also thank C. Regal and D. Jin for sharing their experimental recombination
rates.


\begin{references}

\bibitem{DeMarco} B. DeMarco and D.S. Jin, Science {\bf 285}, 1703 (1999).
\bibitem{OHara}   K.M. O'Hara {\it et al.}, Phys. Rev. Lett. {\bf 85},
                  160407 (2000).
\bibitem{Truscott} A.G. Truscott {\it et al.}, Science {\bf 291}, 2750 (2001).
\bibitem{Fedichev} P.O. Fedichev {\it et al.}, Phys. Rev. Lett. 
                   {\bf 83}, 2921 (1996).
\bibitem{Nielsen} E. Nielsen and J.H. Macek, Phys. Rev. Lett. {\bf 83},
                  1566 (1999).
\bibitem{Esry}    B.D. Esry {\it et al.}, Phys. Rev. Lett. {\bf 83},
                  1751 (1999).  Note that the cross section formula in this
                  reference, and the calculated rates, should be decreased 
                  by a factor of six.
\bibitem{Bedaque} P.F. Bedaque {\it et al.}, Phys. Rev. Lett. 
                  {\bf 85}, 908 (2000).
\bibitem{Braaten} E. Braaten and H.-W. Hammer, Phys. Rev. Lett. 
                  {\bf 87}, 908 (2001).
\bibitem{Bohn} J.L. Bohn, Phys. Rev. A {\bf 61}, 053409 (2000).
\bibitem{Loftus} T. Loftus {\it et al.}, Phys. Rev. Lett. {\bf 88}, 173201 (2002).
\bibitem{Inouye} For example, S. Inouye {\it et al.}, Nature (London) 
                 {\bf 392}, 151 (1998).
\bibitem{ThresholdPaper} B.D. Esry {\it et al.}, Phys. Rev. A {\bf 65} 
                         R010705 (2002).
\bibitem{Suno} H. Suno {\it et al.}, Phys. Rev. A {\bf 65}, 042725 (2002).
\bibitem{efim70} V.M. Efimov, Phys. Lett. {\bf B 33} (1970) 563.
\bibitem{Lin} C.D. Lin, Phys. Rep. {\bf 257}, 1 (1995).
\bibitem{BurkeThesis} J.P. Burke, Jr., Ph.D. thesis, University of 
                      Colorado, 1999.
\bibitem{EsryJPB} B.D. Esry, {\it et al.} J. Phys. B {\bf 29}, L51 (1996).

\end{references}
\end{document}